# Screening Genome Sequences for known RNA Genes or Motifs

Daniel Gautheret

Université Paris-Sud, CNRS-UMR8621, Institut de Génétique et Microbiologie, Bâtiment 400, 91405 Orsay Cedex, France. New address: Institute for Integrative Biology of the Cell, CNRS, CEA, Université Paris-Sud, Gif-sur-Yvette, France.

E-mail address: daniel.gautheret@u-psud.fr

## 1. Introduction

In the mid 1980s, as genomic sequences in the Genbank database reached their first million nucleotides, a few intrepid biologists turned their attention towards screening DNA sequences for the presence of known RNA genes (mostly tRNA at this time) and RNA motifs such as ribozymes or specific stem-loop structures. They soon realized this task was harder than finding homologues of protein-coding genes, since non-coding RNAs were defined not only by their primary sequence, but by a combination of sequence and secondary structure.

Thirty years and hundreds of new RNA families later, this view still holds, albeit with wide variations between individual RNA molecules. For instance, some RNAs have sequence signatures that are stronger than those of most protein-coding genes. Fig. 1A shows an alignment of the Escherichia coli tRNAAla and its homologue in the Bacillus subtilis genome. Although the two species are separated by billions of years of evolution, their tRNAs have retained more than 90% identity. Such homologues may be efficiently identified using a sequence alignment program such as Blast [1]. One may argue, however, that tRNAs are very special molecules: they sustain selective pressure at many sequence positions due to their multiple and universal functions. But experience has shown that a variety of RNA genes, including fast evolving ones such as microRNAs, are also well characterized at the sequence level and can be efficiently detected using sequence comparison programs [2]. Yet, for many other RNAs, primary sequence information is not sufficient for accurate detection. Certain RNA genes such as the RNase

P RNA present little if any conserved sequence, and most RNA structural motifs such as GNRA and E-loops, SECIS elements, U-turns or Rho-independent transcription terminators are composed of a few key nucleotides associated with a larger secondary structure scaffold. Fig. 1B shows consensus structures of the loop E [3] and U-turn [4] motifs, whose functions are conferred by a small set of conserved nucleotides embedded into a larger stem-loop structure. Such motifs cannot be identified using sequence comparison and require specialized programs that take into account both secondary structure and sequence constraints. Mining genomic sequences for RNA elements may thus entail using very different approaches depending on the type of element under scrutiny.

Freyult et al. [5] published a detailed comparison of the performance of various RNA detection programs using simulated sequence sets, and Menzel et al. [2] proposed a more qualitative evaluation of three programs based on “real-life” sequence sets. Here our purpose is not to evaluate programs but instead to help readers choose the most appropriate program for different situations and present practical execution examples for each of the programs. This article updates and expands our previous one in this series [6]. The proposed tasks are accessible to any junior bioinformatician or biologists with a basic knowledge of Unix commands. We provide links to official download sites for the various bioinformatics programs at the end of the Chapter. We assume that all the required programs are properly installed by a system administrator on a computer running the Linux or another Unix operating system.

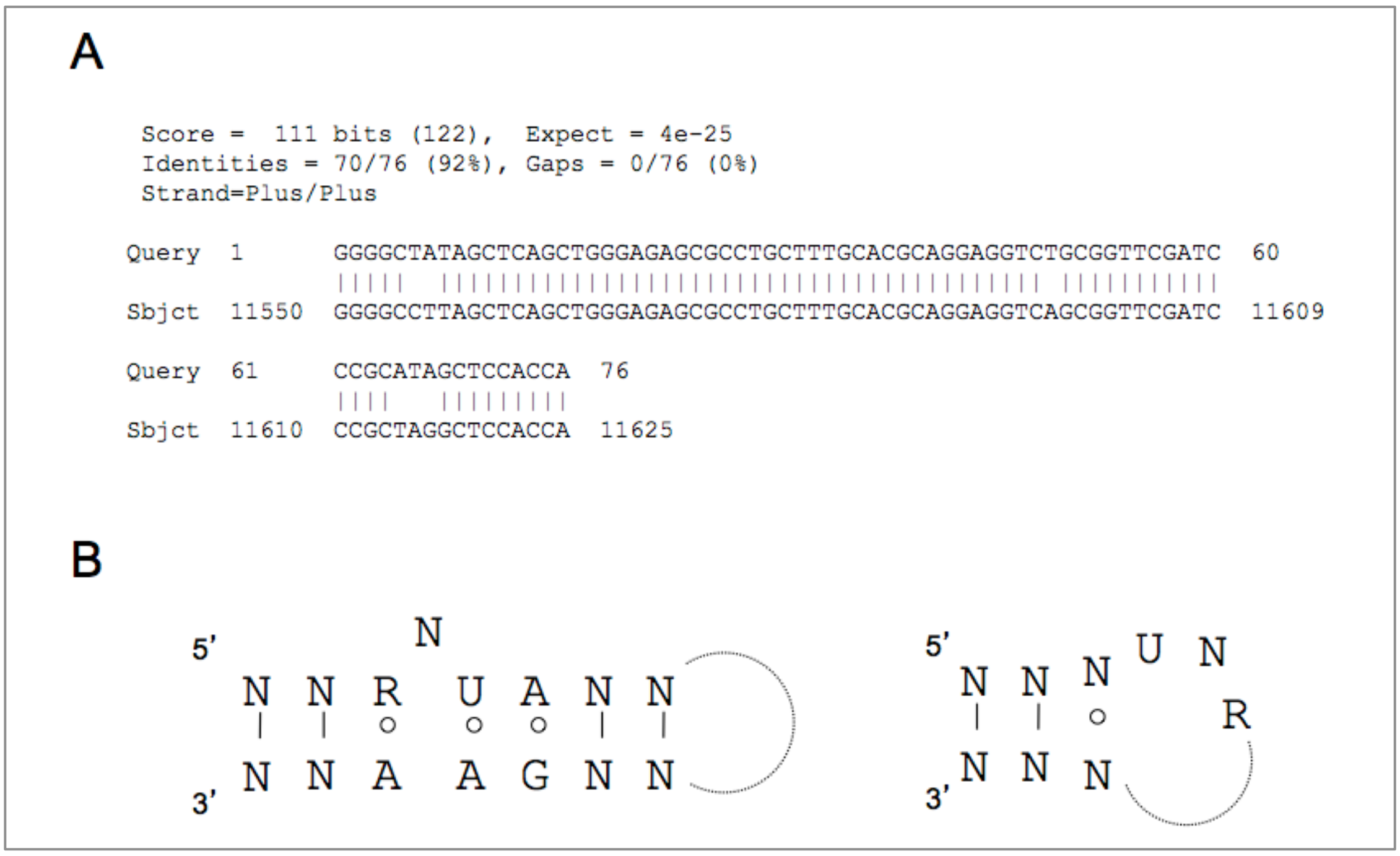

**Fig. 1**. RNA motifs vary widely in their evolutionary constraints. A. A strong primary sequence constraint is visible is this Blast alignment of tRNAAla from E. coli (top) and B. subtilis (bottom). B. Predominant secondary structure constraints are visible in the consensus loop E and U-turn motifs. Thin vertical lines indicate Watson-Crick base pairs, open circles indicate non-canonical base pairs.

# 2. Choosing the Right Search Program

RNA search programs can be roughly divided in two classes: descriptor-based and homology-based. Descriptor-based programs allow biologists to describe, using a defined syntax, the sequence and structure of an RNA motif and search this motif in genome databases. The most widely used program in this family is RNAMOTIF [7]. Homology-based programs constitute a very diverse family. The most popular of them is Blast [1], which uses a single sequence as input and identifies segments of similarity in a sequence database. The HMMER program [8] uses a sequence alignment as input and, consequently, is able to exploit variation and conservation in the gene or motif. The last category of homology search programs, exemplified by ERPIN [9] and INFERNAL [10], use multiple sequences and a description of secondary structure to perform the search. Knowledge of secondary structure allows these programs to identify parts of an RNA motif that are completely variable in sequence while retaining a constant structure. In Table 1, we present an overview of the qualities and domains of application of the different programs.

**Table 1.** Domains of application of different RNA search programs

| Program | Good for | Not good for | Pros/cons |
| --- | --- | --- | --- |
| Descriptor-based (RNAMOTIF) | Short motifs (terminators, hairpin loops, internal loops etc.) | Complete RNA genes | Quick / Requires learning descriptor language |
| BLAST | Long or short RNA genes, domains of genes | Short motifs, distant homologues | Quick and easy / boundaries not reliable |
| HMMER | Long or short RNA genes, domains of genes | Short motifs, distant homologues | Quick / a bit more complex than Blast |
| ERPIN | RNA motifs or genes < 200 nt, domains of RNA genes, RNA with important pseudoknot | Long RNA genes, highly variable structure, irregular stems | Quick / somewhat tricky to use |
| INFERNAL | RNA motifs or genes < 200 nt, distantly related RNAs | Long RNA genes, highly variable structure, pseudoknots | Most sensitive method / Slow |

# 3. Overview of the RNA Search Procedure

Any serious RNA search should address the question of secondary structure. Knowledge of an RNA secondary structure opens the way to the most sophisticated and sensitive search programs. It may also provide important clues about the function of an RNA. If a conserved structure is present in the RNA under study, then it must be identified. There are well-known programs for predicting an RNA secondary structure from sequence.

However, it is believed that on average about 30% of the base pairs predicted by these programs are incorrect. This false prediction rate is definitely too high if one wants to discover further RNAs based on this information. Much more reliable structure prediction can be achieved using a set of homologous RNAs and seeking a secondary structure that is compatible with all sequences in the set. A typical RNA search begins with a step of sequence collection to create this first training set. This can be done using a purely sequence-based program such as Blast. Once this first set is completed, a common RNA structure is derived using a program that simultaneously folds and aligns the sequences in the set. Until recently, such algorithms were considered as too computationally expansive to be applicable to real-life situations and biologists were reduced to align RNAs using a sequence-based multiple-aligner such as CLUSTALW [11], which often yield poorly aligned structures. However, recent progress allows reasonably accurate simultaneous alignment and folding. We chose the LOCARNA [12] program that presents the advantage of running on a web server.

Once a structured alignment is available, it needs to be carefully inspected. Are we sure all RNAs in the set are homologues? Are they complete? Are there obvious misalignments? Some manual curation may be helpful at this stage. Then, one should identify the most conserved sequence and structure elements. This consideration dictates which program can be used to seek further homologues. If no obvious conserved secondary structure stands out, we recommend pursuing with further rounds of BLASTN and the HMMER program. If short sequence/structure motifs appear, then one may use RNAMOTIF to seek further instances of these motifs. If a conserved structure including several stems and loops is visible, then one might resort to more sophisticated programs that will take these structures into account. ERPIN is useful when a core of regular stems is present. INFERNAL is preferred when structures are more variable. The latter is the most general and powerful of all programs used in this Chapter, but it is also the slowest, although recent developments have brought its search speed closer to that of other programs.

As further RNAs of the family are discovered using any of the above tools, they can be used to enrich and improve the structural alignment so that further search iterations can be performed. The steps of a typical RNA search can be summarized as in Fig. 2. All search programs accept sequence databases in the Multifasta format. Recent versions have practically no limitation in the size or number of sequences in the input file.

# 4. Assessing Search Specificity

Most sequence and motif search programs compute a score and an expectation value (E-value) for each hit. An E-value tells us how many hits could be obtained by chance with the same score or higher in a random database of same size. An E-value above 1 or even 0.1 is likely to occur by chance. Any E-value below $10^{-2}$ should in principle convince peer scientists that a biologically related motif has been found. However, keep in mind that a low E-value does not demonstrate biological significance, as it is subject to artifacts caused for instance by low complexity regions in genomes. Conversely, hits with a poor E-value may turn out to be true homologues that have diverged a lot. Also note that E-values depend on the database size. When a search program is run repeatedly on separate sequence files (e.g. with one file per chromosome), the E-value is valid only for each distinct file.

RNAMOTIF is the only program in our protocol that does not automatically compute E-values. In this case, or when E-values are not trusted for some reason, it is common to use an expected density of False Positives (e.g. FP per Mb) as a measure of specificity. To this aim, we may run the search against a randomized sequence database. Random sequences with a uniform nucleotide composition (25% each) are not advised, since background compositions (single-nucleotide and di-nucleotide frequencies) strongly affect the number of chance hits: for instance, an AU-rich motif is more likely to occur in an AU-rich sequence background. It is therefore important to reproduce in the negative control set the overall compositional biases of the target database. Several programs can do this. We will use *shuffle*, by S. Eddy (see Section Program Versions and Download Sites), that randomizes a sequence while preserving the single- and dinucleotide frequencies.

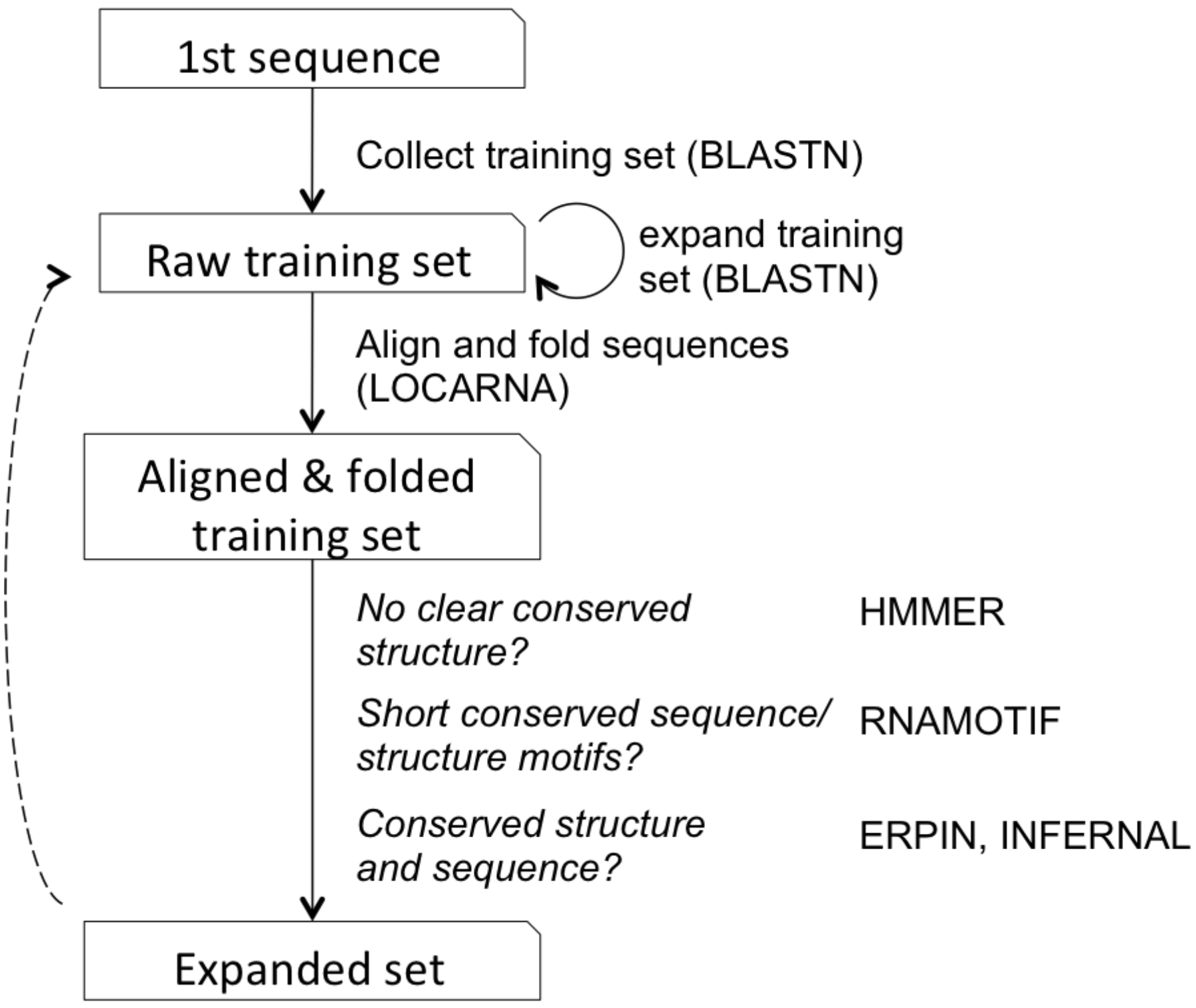

**Fig. 2.** Outline of a typical RNA motif search protocol. The protocol starts with a BLASTN sequence collection followed by structural alignment (align & fold) and the use of a specialized search software.

The randomization procedure can be performed on the target database, so that each genome in the database is shuffled using its own compositional characteristics. When the search database is not decided beforehand, an alternative option is to shuffle the training set itself. This is a very conservative approach, as a randomized training set, owing to its limited sequence space, is more likely to give rise to false positives than any other sequence.

If specificity has to be quantified, the negative set should, ideally, be large enough to find at least a few False Positives. By default, the *shuffle* program produces a sequence of same size as the input sequence. When using a short sequence such as the training set, the shuffling procedure must be iterated to produce a random sequence of the desired size. For instance:

```
shuffle -d -n 100 trset.fasta > trset.rnd
```

Here the sequence to be randomized is in file `trset.fasta` (multi-fasta format) and the output is redirected to `trset.rnd`. The `-d` option is used to preserve dinucleotide composition in the random sequence, and `-n 100` specifies 100 shuffling iterations. The randomized sequences from each iteration round are concatenated and will thus be 100 times larger than the original sequence.

# 5. A Test Case: Looking for Homologues of a Bacterial sRNA

Our test challenge here is to discover homologues of a bacterial small RNA (sRNA) gene named RsaE, a 113-bp RNA first identified in *Staphylococcus aureus* [13, 14]. We will try to stay as close as possible to a "real life" situation, which means we assume only one instance of the RNA molecule is known first (here the *S. aureus* RsaE), and no structure information is available. This is the situation shown in the first box of Fig. 2. The initial, experimentally confirmed, sequence is entered in the Fasta format in a file named `rsae.fa` (see Section 7, "Supplemental data"), as follows:

```
>Rsae
AUGAAAUUAAUCACAUAACAAACAUACCCCUUUGUUUGAAGUGAAAAAUUUCUCCCAUC
CCCUUUGUUUAGCGUCGUGUAUUCAGACACGACGUUUUUUUAU
```

Our objective is to find all other instances of this sequence in bacterial genomes. Our search database will contain all complete genomes retrieved from the NCBI web site. An initial set of genomes from 1080 species was downloaded in 2010 and reduced to 732 species by retaining a single instance of each species. The sequences are saved in a single multi-Fasta file of about 2.5 Gb named `732bacs.fa` (see Section 7). To simplify command lines, we assume all data files are present in the working directory.

## 5.1 *Building a First Training Set with BLASTN*

Blast is the program of choice for a fast and easy identification of a first set of homologues. Biologists are familiar with the NCBI BLAST web interface. Here we use the command line version of the program. An advantage of using command line is that users can scan "private" sequences not yet included in the NCBI sequence databases. Furthermore, playing with parameters is faster with command lines. We assume a recent version of BLAST (such as 2.2.25) is installed on the reader's workstation.

A local BLAST run requires a properly formatted database. Formatting our database file `732bacs.fa` is done using:

```
makeblastdb -in 732bacs.fa -dbtype 'nucl'
```

This command generates several index files required for BLAST runs. You should not worry about the names of these files, as you will always refer to the Fasta file name (here `732bacs.fa`) for further operations. There are several programs in BLAST. Here we use BLASTN, which is used to scan nucleic acid databases for nucleic acid sequences. With our data files, a simple BLASTN command would look like:

```
blastn -db 732bacs.fa -query rsae-sau.fa
```

The command takes about 0.7 seconds to execute and produces a lot of output. It is pretty difficult to understand how many hits were exactly recorded. A simpler table output can be obtained using the `-outfmt 7` parameter, such as in:

```
blastn -db 732bacs.fa -query rsae.fa -outfmt 7
```

Now we can count 12 hits from 12 different species. Are they all statistically significant though? The lowest scoring hit is from the *Geobacillus* strain WCH70. Its E-value of $3 \times 10^{-11}$ is highly significant. We can control the E-value cutoff using the `-evalue` parameter, for instance by appending `-evalue 0.01` to the BLAST command.

Can we find more distant homologues using BLASTN? This question is commonly addressed by reducing word size. In BLAST, both the database and the query sequences are broken down into k-length words (k-words). BLAST is able to quickly scan these word lists for exact or near-exact matches. Longer alignments with possible gaps and mismatches are generated only after one k-word match is found. This enables a very fast search at the risk of missing some hits that would contain no exact or near-exact k-word match. The default word size for BLASTN is 25. It can be changed using the word_size parameter. Shorter words will increase sensitivity, at the cost of longer runtimes. Let's reduce word size to 10:

```
blastn -db 732bacs.fa -query rsae.fa -outfmt 7 -evalue 0.01 –word_size 10
```

The command now takes 2.1 seconds to run, but finds 21 hits from different species, that is nine more than with the default word size. Note how significant these new alignments are. The lowest scoring hit is now found in *Geobacillus kaustophilus*, and has an E-value of $9 \times 10^{-10}$, which is still very significant. Reducing word size to its minimum value (k=5) extends runtime to nearly 1 minute but does not reveal further solutions.

This list of 21 BLAST hits is a good starting point for further analysis. We should now extract the corresponding sequences. The latest version of BLAST provides useful output controls. It is now possible to retrieve automatically the matched sequences and other items using parameter `-outfmt 10` followed by the desired item names. For instance replacing `-outfmt 7` in the above command with `-outfmt "10 evalue sseqid sseq"` will output the E-value, database sequence Id and matching sequence region in the database for each hit.

A visual inspection of these sequences reveals that several species share the exact same RsaE sequence. After discarding the redundant hits, our raw training set is composed of 14 distinct sequences. Before going any further, we should pay attention to the species distribution of these sequences. We have found RsaE in *Staphylococcus*, *Macrococcus*, *Bacillus*, *Geobacillus*, *Anoxybacillus* and *Lysinibacillus*. These genera all belong to the "Bacilli" class of the Firmicute phylum. This is an important observation that supports the relevance of the hits. Hits distributed in multiple distant phyla would be a warning sign of possible false positives.

### *5.2 Alignment and Structure Prediction*

As explained above, establishing the RNA secondary structure is an important step for further homology search. To this aim, we will use the LOCARNA program [12] through its Freiburg web server. LOCARNA belongs to a family of programs that simultaneously align and fold a set of RNA sequences. These programs are known to provide better RNA alignments than those relying only on sequence. The LOCARNA server is intuitive and works fine with default parameters. We can just paste the 14 Fasta-formatted sequences into the input window. To produce a Fasta file from the BLAST output, any text editor will do. As LOCARNA does not accept gap characters, we will use the editor to remove gaps too. The Fasta-formatted raw training set is available as Suppl. file `trset.fa` (see Section 7). The LOCARNA server displays the structural alignment after a few minutes. The output is shown in Fig. 3A with parentheses representing the secondary structure. This predicted structure is remarkably similar to the one established for *S. aureus* using chemical and enzymatic probing [13] (Fig. 3B).

LOCARNA, like any other multiple alignment program, does not produce perfect alignments. For instance, the AAAC in the 5' region of *G. kaustophilus* RsaE would probably be aligned better with the AAAC segments found in all other sequences. To keep things simple, we will not try to improve this alignment manually, and we will proceed with the one in Fig. 3A.

It appears from our alignment that RsaE homologues differ in their 3' regions.  Either some RNAs are actually shorter, or this may be an effect of the BLAST program failing to extend the alignment to its actual ending. To settle this, it is necessary to retrieve the corresponding regions from the bacterial genomes and observe their sequence. This can be done using the coordinates from the BLAST output and any genome browser such as the Ensembl Bacterial Genome Browser (http://bacteria.ensembl.org/). We retrieved the missing 3' regions and realigned the sequences using LOCARNA. These 3' regions differed strongly from other RsaE sequences and aligned poorly both at the sequence and structure level (not shown). Therefore we can assume that these 3' regions are either missing or too different from the original RsaE sequence to be kept in our alignment. We also noted that the *G. kaustophilus* sequence lacked about 15 nt in its 5' region. We retrieved the corresponding segment from the genome sequence and used it for the LOCARNA alignment. The extended sequence aligned nicely with other RsaE sequences although parts of the long 5' stem appeared not to be conserved. The extra segment (italics) was incorporated into Fig. 3A.

The RsaE initial alignment presents a long stem on the 5' side. However, the sequence of this stem is highly conserved and there is little evidence of covariation that would indicate the base pairs are under evolutionary pressure. Therefore we cannot be absolutely sure about this secondary structure. The 3' part, which is missing in half the sequences is a typical Rho-independent transcription terminator, with a stem loop followed by a U-rich region. We may hypothesize that the RsaE sequences lacking this terminator use another termination mechanism.

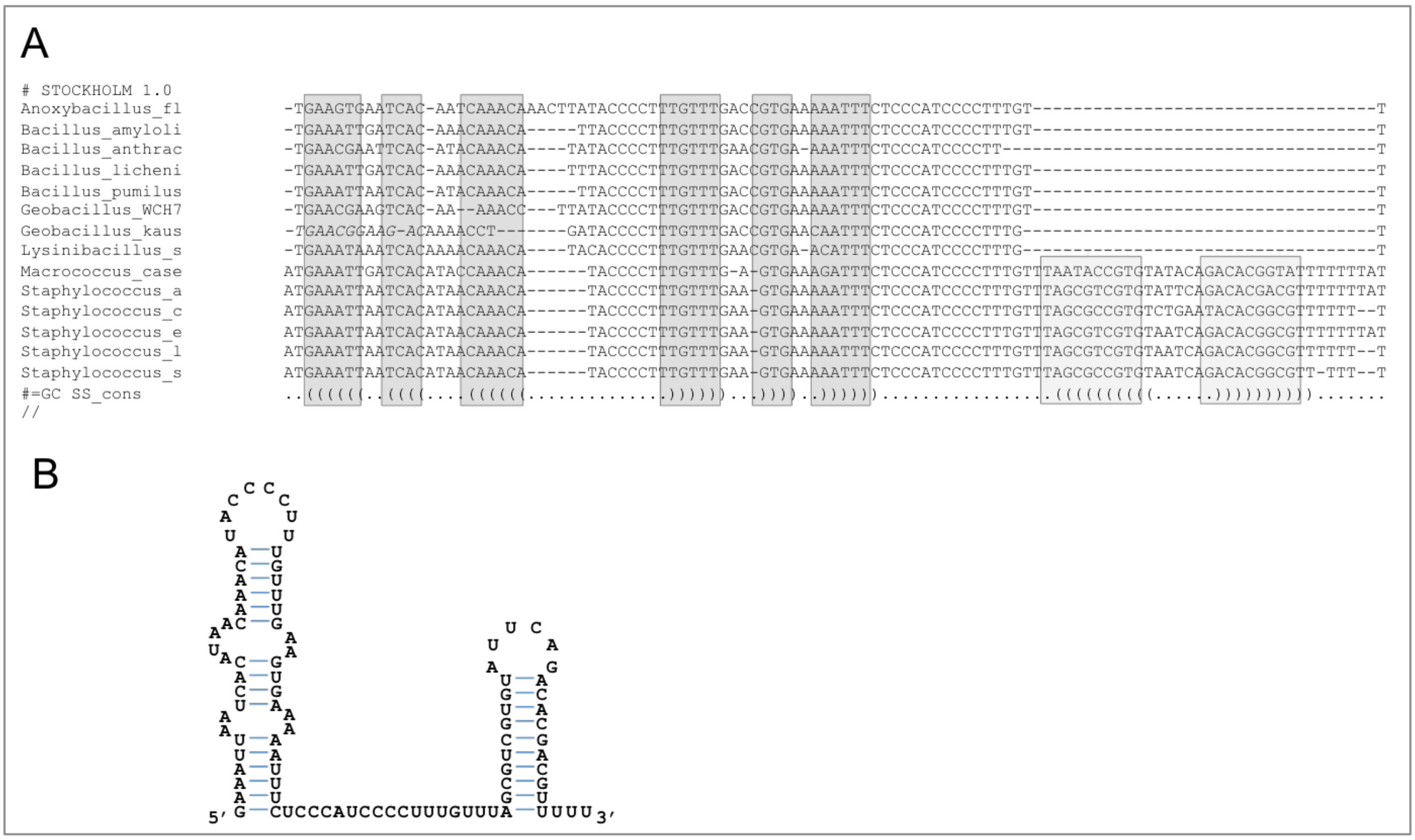

**Fig. 3. A.** An aligned and folded training set of RsaE sequences (Stockholm format) obtained from the initial BLAST screen and aligned using LOCARNA. Parentheses in the bottom line indicate base-paired positions, highlighted by shaded boxes. Bases in italics in "Geobacillus_kaus" were absent in the initial BLAST hit and manually inserted before alignment. **B.** The *S. aureus* RsaE secondary structure derived from enzymatic and chemical probing [13].

## *5.3 SEARCHING WITH HMMER*

A sequence alignment is a powerful means of conducting homology searches, even in the absence of secondary structure information. The main reason is that an alignment permits to distinguish variable regions that are not crucial for RNA function from conserved, crucial regions, and thus enables weighting matches accordingly. A good way to exploit an alignment is to use the HMMER package [8]. HMMER converts the alignment into a profile Markov model that records both the statistics of base frequencies in each alignment column as well as that of consecutive bases in neighbouring columns. Once a profile HMM model is built from the alignments, matches to this model in sequence databases can be sought using the dynamic programming algorithm in HMMER. First, the training alignment should be formatted in the Stockholm format, as shown in suppl. file `trset.stk`. Here is the command to build the profile HMM:

```
hmmbuild trset.hmm trset.stk
```

This creates a file named `trset.hmm`, which can be used for scanning genome sequences. The search command is then entered as follows:

```
hmmsearch trset.hmm 732bacs.fa
```

After a runtime of 42 seconds, we obtain 18 solutions. These include new Bacillus and Geobacillus species, as well as an entire new genus, Oceanobacillus, that was absent in the initial training set. This HMMER hit is shown in Fig. 4. The figure legend explains the different output items. The default mode of hmmsearch uses several heuristics to speed up the search, at a certain cost for search sensitivity. It is possible to switch off the heuristics using the –-max option. This slows down the search by a factor of 10, but reveals no novel significant hit in our case.

```
>> Oceanobacillus_iheyensis_BA000028
   #    score  bias  c-Evalue  i-Evalue hmmfrom  hmm to    alifrom  ali to    envfrom  env to     acc
 ---   ------ ----- --------- --------- ------- -------    ------- -------    ------- -------    ----
   1 !   22.5   0.1   3.2e-07   1.4e-05      15      67 .. 1267728 1267782 .. 1267716 1267784 .. 0.80

  Alignments for each domain:
  == domain 1    score: 22.5 bits;  conditional E-value: 3.2e-07
                                      ...((((((.-.........))))))..-)))...))))))............... CS
                           trset1      15 aaacaaaca.auaccccu.uuguuugaacgugaaaaauuucucccaucccc.uuug 67
                                      a acaaaca auaccccu uuguuuga c ug aaa uuucucccaucccc uuug
  Oceanobacillus_iheyensis_BA000028 1267728 AUACAAACAaAUACCCCUuUUGUUUGAUCAUG-AAACUUUCUCCCAUCCCCcUUUG 1267782
                                      567888888889999996268***********.5678*********987515555 PP
```

**Fig. 4.** Part of an HMMER output. The local hit is shown with 5 lines. The “CS” line is just a reminder of the secondary structure in the .stk file, which is not used by HMMER, the next three lines show the consensus profile, matching residues and target sequence, and the last line shows the posterior probabilities of each aligned residue which is an indication of alignment reliability. Values range between 0 and *, with 0 corresponding to a 0-5% probability, up to * for a 95-100% probability that this residue was produced from the hidden markov model created from the input alignment.

### *5.4 Searching with RNAMOTIF*

The RNAMOTIF program permits to describe an RNA motif, including sequence and secondary structure constraints, and to find all instances of this motif in a sequence database. Therefore, a major difference from the previous approaches is that we can now use secondary structure information. A typical RNAMOTIF descriptor is shown in Fig. 5A. Let’s take a quick look at this language. The first section (parms), says that G:U pairs are allowed in addition to Watson-Crick pairs. The next section (descr) is the actual descriptor, listing all secondary structure elements. There are three types of elements: h5 stands for the 5’ strand of a helix, h3 for the 3’ strand of a helix, and ss for a single strand. When no other indication is provided, RNAMOTIF assumes pairing between h3 and h5 elements in a purely nested fashion, just as in a parenthesis notation. Non-nested helices (*i.e.* pseudoknots) require a special tag that modifies this default behavior. Associated to each h5, h3 or ss element is a list of optional parameters that specify minimum and maximum length, conserved sequences, conserved base pairs, etc. By default, RNAMOTIF reports all sequences matching the descriptor. Sometimes, however, additional filters are required that cannot be applied at the "descr" level. An optional score section is used for this purpose. In this section, tests can be performed using a C-like computer language, and matches are rejected if the tests fail. For an in-depth explanation of RNAMOTIF descriptors and scoring functions, refer to the detailed manual included in the program.

A

```
parms
   wc+=gu;

descr
  ss(len=1, seq="U")
  h5(len=6, mispair=1, ends='mm')
    ss(len=2)
    h5(len=4, mispair=1, ends='mm')
      ss(minlen=2, maxlen=4, seq="^A")
      h5(len=6, mispair=1, ends='mm')
        ss(minlen=7, maxlen=14, seq="ACCCCUU$")
      h3
      ss(minlen=1, maxlen=3, seq="^A")
    h3
    ss(minlen=1, maxlen=3, seq="^AA")
  h3
```

B

```
descr
  ss(len=60, seq="^.............AAAC.\{2,10\}ACCCCUUUGUUUGA")
```

C

```
descr
  ss(len=20, seq="AAAC$")
  ss(minlen=2, maxlen=10)
  ss(len=14, seq="ACCCCUUUGUUUGA", mismatch=1)
```

**Fig. 5.** RNAMOTIF descriptors for RsaE. **A.** This descriptor uses three stems with variable sequences and six single strands containing conserved sequences. **B.** This descriptor uses a single 60-nt single strand with a long conserved 3' motif. **C.** This descriptor splits the previous single strand in three parts. In the last part, a mismatch is allowed in the consensus motif.

Let's see how our knowledge of the RsaE RNA can be turned into an efficient descriptor. We will use only the most basic features of the RNAMOTIF language. To build our first descriptor (Fig. 5A), we make the typical assumption that a regulatory RNA has variable sequences in the secondary structure helices and conserved sequences in single-stranded regions, as the latter are supposed to undergo tertiary interactions. This is of course a simplistic view, but we would like to see how it performs. We set helix lengths as in the Fig. 3 consensus (size = 6, 4 and 6 nt, respectively from 5' to 3'). To account for shorter helices, we allow 1 mismatch per helix (`mispair=1`) and, since the default behavior of this parameter restricts mispairs to internal positions, we use `ends='mm'` to allow for mispairs at helix ends. We specify conserved sequences in several single-stranded segments. The most conserved single-stranded region is the apical loop where the conserved seven 3'-proximal bases are specified by `seq="ACCCCUU$"`. In the fifth "ss" region `seq="^A"` means that this strand should start with an A. We save this descriptor file as `descriptor.txt` and we are now ready to look for instances of this motif in our database. The basic RNAMOTIF command then looks like:

```
rnamotif -descr descriptor.txt 732bacs.fa
```

The run takes about 1 minute and produces a pretty large output. It is better to use a redirection to store hits in an output file. Also, due to variations in helix sizes and positions, RNAMOTIF tends to generate several overlapping solutions for each actual motif. The utility program `rmprune` is provided to filter out these overlapping solutions and retain only the highest scoring one at each site. This program is executed in combination with RNAMOTIF using the "pipe" sign (|) as follows:

```
rnamotif -descr descriptor.txt 732bacs.fa |rmprune > outfile
```

From now on, we will assume that every RNAMOTIF run is performed using `|rmprune > outfile`. We can now count 16 solutions produced by the RNAMOTIF run. Sample outputs are shown in Fig. 6. Most hits are from the genera *Bacillus*, *Macrococcus*, *Staphylococcus*, and *Anoxybacillus*. A few training set sequences such as the Geobacilli are missing, which indicates our descriptor is too strict in some respects. Indeed, we can see in the alignment that shorter helices #2 and #3 in *G. kaustophilus* do not match the descriptor constraints.

**A**

```
>Anaerococcus_prevotii_DSM_20548_CP001708
Anaerococcus_prevotii_DSM_20548_CP001708    0.000 0  212356   74
t ttcttc aa ccaa at ccaaaa caactacccctt ctttgg agt tttg aa gagaaa
```

**B**

```
>Streptococcus_uberis_0140J_AM946015
Streptococcus_uberis_0140J_AM946015    0.000 0  430954   72
tgctactgatggtatcaaac cagactta accttttgtttga tgtaccttcagaaattggcatagcgcgcat

>Clostridium_phytofermentans_ISDg_CP000885
Clostridium_phytofermentans_ISDg_CP000885    0.000 0 2755278   69
gcatcctggttttacaaaac tacct acccctttttttga atatattacaccctctaattctaattgcgc
```

**Fig. 6.** Extracts of RNAMOTIF outputs. Hits are formatted as follows. Line 1: title line of Fasta file entry. Line 2: Sequence name, score (here our descriptors do not include a scoring section hence all scores are 0), strand ("0" indicates plus strand and "1" minus strand), hit position, hit length in nt. Line 3: hit sequence, with stems and single-stranded elements separated by spaces. **A.** A hit from the descriptor in Fig. 5A. **B.** Two hits from the descriptor in Fig. 5C.

Interestingly, only four sequences are from species outside our training set. Considering the large size of the database, this is a small number. These include one hit from *Anaerococcus prevotii* (Fig. 6A). Albeit distant in the phylogenetic tree, this species belongs to the Firmicutes, as do the other known RsaE sequences, and is therefore an interesting candidate to consider. However, we observed that the sequence just following the motif is not as CU-rich as in other RsaE sequences. The other new hits are from phyla that are very distant from Firmicutes and therefore unlikely to contain an RsaE homologue. We can reasonably assume they are false positives.

To assess the significance of new RNAMOTIF hits, we advise using a shuffled database as a control. Here we will randomize our bacterial genome database while preserving the dinucleotide frequencies, using the `shuffle` command as this:

```
shuffle -d 732bacs.fa > 732bacs.rnd
```

Running the same RNAMOTIF descriptor on the shuffled database `732bacs.rnd` produces 6 hits. This is a clear warning that the few hits obtained with real data can be false positives.

Examination of the RsaE alignment suggests that sequence alone may describe RsaE better than secondary structure. The stems are indeed highly conserved in sequence and show no sequence covariation to support their base-pairing potential. Let's see how efficient our RsaE search could be by using solely primary sequence constraints. All RsaE members harbor a constant AAAC motif followed by a highly conserved CU-rich sequence that is repeated in tandem. The descriptor in Fig. 5B describes one copy of the tandem motif in a single line. The syntax for sequence constraints in RNAMOTIF is that of Unix regular expressions. The "^" sign indicates the beginning of the element and the string `.\{2,10\}` describes a sequence of 2 to 10 unspecified bases, described by the dot character. This simple motif performs surprisingly well. After 75 seconds, RNAMOTIF identifies 27 hits, covering all the species in the training set plus several other species from the same genera and one hit in *Pectobacterium wasabiae*, an enterobacterium. Enterobacteria are very distant from Firmicutes, which makes this hit an unlikely RsaE candidate.

Running the descriptor in Fig. 5B against the shuffled genome database produces a single hit, which shows that our simple sequence motif is more specific than the complex secondary structure motif in Fig. 5A. Can we now relax this descriptor and find additional candidates? RNAMOTIF allows for mismatches in sequence motifs, using the `mismatch=n` parameter (n being the number of allowed mismatches). However, the mismatch parameter is compatible only with regular expressions of constant length. Therefore we cannot allow for mismatches in a variable region of size 2 to 10 as above. We thus have to rewrite our descriptor using distinct `ss` elements, as shown in Fig. 5C.

The descriptor in Fig. 5C that tolerates one mismatch in the CU-rich region takes longer to run (2 minutes) than the exact descriptor and produces 71 solutions. In addition to new species from the same genera as above, these include a number of apparent false positives from Cyanobacteria or Enterobacteria, plus new Firmicutes candidates of the genera *Streptococcus* and *Clostridium* (Fig. 6B). These results should be considered with caution though, as the false positive ratio has increased significantly. We now find 35 hits in the randomized database (not shown).

A possible control to evaluate the new candidates is to analyze the sequence downstream of the motif. We can see in the training set (Fig. 3) that the downstream sequence is a near exact repeat of the central conserved region. Flanking sequences around a given motif are easy to extract using RNAMOTIF by adding new “ss” elements of desired length with no sequence constraint. We used this to obtain the 30 nt of downstream sequences. This sequence is not UC-rich in Streptococci, but it is in one of the Clostridia, *C. phytofermentans* (Fig. 6B). Therefore this candidate may be considered further.

We can see that RNAMOTIF is a valuable tool to test ideas about what determines an RNA motif. We have developed highly specific descriptors in just minutes, using either primary or secondary structure information. One important limitation of RNAMOTIF though is its lack of an objective scoring function. Although the program accepts user-defined scoring functions, their quality will depend on the user’s appreciation of the RNA motif. The subsequent protocols address this issue.

### 5.5 SEARCHING WITH ERPIN

ERPIN circumvents the descriptor design problem by extracting information directly from a structure-annotated alignment and building log-odd score profiles for each stem and single-stranded segment in the alignment.

A

```
>#=GC SS_cons
000000000000000000000000000000000000000000011110010110000001111111111111111111111111111111111111111111111111
112222223345667777788999999999999999999998811116634552222227777777777777788888888888888888888888888888888888
>Anoxybacillus_fl
-TGAAGTGAATCAC-AATCAAACAAACTTATACCCCTTTGTTTGACCGTGAAAAATTTCTCCCATCCCCTTTGT----------------------------------T
>Bacillus_anthrac
-TGAACGAATTCAC-ATACAAACA----TATACCCCTTTGTTTGAACGTGA-AAATTTCTCCCATCCCCTT-------------------------------------T
>Geobacillus_WCH7
-TGAACGAAGTCAC-AA--AAACC---TTATACCCCTTTGTTTGACCGTGAAAAATTTCTCCCATCCCCTTTGT----------------------------------T
>Macrococcus_case
ATGAAATTGATCACATACCAAACA------TACCCCTTTGTTTG-A-GTGAAAGATTTCTCCCATCCCCTTTGTTTAATACCGTGTATACAGACACGGTATTTTTTTAT
>Staphylococcus_a
ATGAAATTAATCACATAACAAACA------TACCCCTTTGTTTGAA-GTGAAAAATTTCTCCCATCCCCTTTGTTTAGCGTCGTGTATTCAGACACGACGTTTTTTTAT
../..
```

B

```
     SS HHHHHH SS H S HH SSSSS HH SSSSSSSSSSSSSSSSSSSS HH SSSS HH S H SS HHHHHH SSSSSSSSSSSSSS SSSSSSSSSSSSSSSSSSSSSSSSSSSSSSSSSSS
     00 000000 00 0 0 00 00000 00 00000000000000000000 00 1111 00 1 0 11 000000 11111111111111 11111111111111111111111111111111111
     11 222222 33 4 5 66 77777 88 99999999999999999999 88 1111 66 3 4 55 222222 77777777777777 88888888888888888888888888888888888
     -* ****** ** * - ** -**-- ** ---------*********** ** *-*- ** * * -* ****** ************--- ----------------------------------*

1    -T GAAGTG AA T C AC -AATC AA ACAAACTTATACCCCTTTGT TT GACC GT G A AA AATTTC TCCCATCCCCTTTGT ----------------------------------T
2    -T GAAATT GA T C AC -AAAC AA ACA-----TTACCCCTTTGT TT GACC GT G A AA AATTTC TCCCATCCCCTTTGT ----------------------------------T
3    -T GAACGA AT T C AC -ATAC AA ACA----TATACCCCTTTGT TT GAAC GT G A -A AATTTC TCCCATCCCCTT--- ----------------------------------T
4    -T GAAATT GA T C AC -AAAC AA ACA----TTTACCCCTTTGT TT GACC GT G A AA AATTTC TCCCATCCCCTTTGT ----------------------------------T
5    -T GAAATT AA T C AC -ATAC AA ACA-----TTACCCCTTTGT TT GACC GT G A AA AATTTC TCCCATCCCCTTTGT ----------------------------------T
6    -T GAACGA AG T C AC -AA-- AA ACC---TTATACCCCTTTGT TT GACC GT G A AA AATTTC TCCCATCCCCTTTGT ----------------------------------T
7    -T GAACGG AA G - AC AAAAC CT -------GATACCCCTTTGT TT GACC GT G A AC AATTTC TCCCATCCCCTTTG- ----------------------------------T
8    -T GAAATA AA T C AC AAAAC AA ACA----TACACCCCTTTGT TT GAAC GT G A -A CATTTC TCCCATCCCCTTTG- ----------------------------------T
9    AT GAAATT GA T C AC ATACC AA ACA------TACCCCTTTGT TT G-A- GT G A AA GATTTC TCCCATCCCCTTTGT TTAATACCGTGTATACAGACACGGTATTTTTTTAT
10   AT GAAATT AA T C AC ATAAC AA ACA------TACCCCTTTGT TT GAA- GT G A AA AATTTC TCCCATCCCCTTTGT TTAGCGTCGTGTATTCAGACACGACGTTTTTTTAT
11   AT GAAATT AA T C AC ATAAC AA ACA------TACCCCTTTGT TT GAA- GT G A AA AATTTC TCCCATCCCCTTTGT TTAGCGCCGTGTCTGAATACACGGCGTTTTTT--T
12   AT GAAATT AA T C AC ATAAC AA ACA------TACCCCTTTGT TT GAA- GT G A AA AATTTC TCCCATCCCCTTTGT TTAGCGTCGTGTAATCAGACACGGCGTTTTTTTAT
13   AT GAAATT AA T C AC ATAAC AA ACA------TACCCCTTTGT TT GAA- GT G A AA AATTTC TCCCATCCCCTTTGT TTAGCGTCGTGTAATCAGACACGGCGTTTTTT--T
14   AT GAAATT AA T C AC ATAAC AA ACA------TACCCCTTTGT TT GAA- GT G A AA AATTTC TCCCATCCCCTTTGT TTAGCGCCGTGTAATCAGACACGGCGTT-TTT--T
```

**Fig. 7.** ERPIN data formats. **A.** Typical ERPIN input file (.epn file) showing secondary structure (first three lines) and aligned sequences (next lines). The format is explained in text. **B.** A .epn file displayed using the tview command. Each secondary structure element, S (single-stranded) or H (helical), is highlighted with element number (read vertically) and sequence conservation (asterisk).

First, we need to convert the alignment into the ERPIN format. Fig. 7A shows how the first lines of our RsaE alignment should look like. The first three lines provide secondary structure information. In this example, the secondary structure is the same as the parenthesized structure in Fig. 3, but each structural element (single-stranded or helical) is assigned a different number that is read vertically. When distant segments have the same number (e.g. 02 or 04 in this example), the corresponding element is understood to be helical, otherwise it is single-stranded. The secondary structure lines can be created manually using a text editor, or automatically from a parenthesized secondary structure file using the `parent2epn.pl` script provided in the ERPIN distribution. The lines following the structure information contain the aligned sequences in the classical FASTA format.

The initial parenthesized secondary structure had a single long single-stranded segment composed of elements 17 and 18. Element 18 (the terminator) is not present in all sequences while element 17 is present everywhere and is highly conserved. To enable searching for the conserved part while ignoring the other, it is preferable to split this single-stranded segment into two different elements. We thus created element 18 using a text editor. Our ERPIN formatted alignment was saved as `trset.epn` (suppl. data).

The `tview` command is useful to visualize constraints in the sequence alignment:

```
tview trset.epn
```

The result is shown in Fig. 7B. ERPIN users must decide which part of the alignment is used for the search. A direct search for a multiple-element region could be prohibitive in terms of CPU and memory usage, especially when elements vary a lot in size. Here, the alignment suggests that the region from the 5' side of helix 02 to single strand 17 is the most conserved (asterisks on top). We will thus launch the search using this region (we assume all ERPIN commands below are complemented by `-silent > outfile` in order to send output to file `outfile`, devoid of trace text).

```
erpin trset.epn 732bacs.fa -2,17 -nomask
```

The minus sign before 2 is used to specify the 5' part of stem 02. The `-nomask` argument indicates that all elements in the region should be used for the search. We will see later how we can mask/unmask elements. The ERPIN command takes as much as 4 days to be executed, producing 21 solutions. A sample is shown in Fig. 8A. The solutions cover all species from the training set, plus a few closely related species. The search is highly specific, with all E-values around $10^{-20}$, however it takes a long time and does not find distant homologues.

Using the whole structure for search is thus not very practical here (4 days run!), and it can become worse for RNA motifs containing more elements, more gaps or fewer conserved regions than the RsaE RNA. More gaps in single-stranded segments (helices cannot contain gaps) and more variation in the alignment means longer search times and reduced search specificity. One way to reduce runtime is to restrict the search to a smaller conserved region. For instance we can use the region from single strand 9 to single strand 17:

```
erpin trset.epn 732bacs.fa 9,17 -nomask
```

The search time is now down to a more reasonable 10 hours and ERPIN finds the same 21 hits as before. However, because of the different search region, the hit sequences and E-values are different (Fig. 8B).

A

```
>Anoxybacillus_flavithermus_WK1_CP000922
RC   1 2094941..2095011  108.53  1.23e-22
GAAGTG.AA.T.C.AC.-AATC.AA.ACAAACTTATACCCCTTTGT.TT.GACC.GT.G.A.AA.AATTTC.TCCCATCCCCTTTGT
>Bacillus_amyloliquefaciens_FZB42_CP000560
FW   1 1134164..1134229  115.95  5.80e-26
GAAATT.GA.T.C.AC.-AAAC.AA.ACA----T-TACCCCTTTGT.TT.GACC.GT.G.A.AA.AATTTC.TCCCATCCCCTTTGT
>Bacillus_anthracis_str_Sterne_AE017225
FW   1 1170862..1170925  100.75  3.59e-20
GAACGA.AT.T.C.AC.-ATAC.AA.ACA----TATACCCCTTTGT.TT.GAAC.GT.G.A.-A.AATTTC.TCCCATCCCCTTT--
```

B

```
>Anoxybacillus_flavithermus_WK1_CP000922
RC   1 2094941..2094993  88.48  2.65e-16
ACAAACTTATACCCCTTTGT.TT.GACC.GT.G.A.AA.AATTTC.TCCCATCCCCTTTGT
>Bacillus_amyloliquefaciens_FZB42_CP000560
FW   1 1134182..1134229  90.94  2.15e-17
ACA----T-TACCCCTTTGT.TT.GACC.GT.G.A.AA.AATTTC.TCCCATCCCCTTTGT
>Bacillus_anthracis_str_Sterne_AE017225
FW   1 1170880..1170925  82.78  3.41e-14
ACA----TATACCCCTTTGT.TT.GAAC.GT.G.A.-A.AATTTC.TCCCATCCCCTTT--
```

C

```
>Oceanobacillus_iheyensis_BA000028
FW   1 1267734..1267781  43.42  2.82e+00
ACA----AATACCCCTTTTG.tttgatcatgaaactttc.......TCCCATCCCCCTT-T
```

**Fig. 8.** Extracts of ERPIN outputs. Hits are formatted as follows. Line 1: title line of Fasta file entry. Line 2: Forward (FW) or Reverse (RC) strand, a fixed tag, hit positions, PWM score and E-value. Line 3: hit sequence. Structural elements are separated by dots. Consecutive dots indicate that corresponding elements were not included in the ERPIN command. Gaps inside elements are used to align the hit sequence onto the training alignment. **A, B, C**: results of different ERPIN runs (see text).

It is relatively easy to further decrease CPU time by implementing a search strategy. The main idea is to perform a stepwise search using masks. In the first step, one should mask most of the selected region and retain only a few key elements. ERPIN will disregard all masked elements and restrict search to the unmasked part. The best parts to unmask during the first step are those conserved elements that occur rarely in the database and are in close proximity to each other so that they are quickly identified. Then, at each successive step, one unmasks additional elements for ERPIN to look for. ERPIN will consider these latter elements only after elements have been identified in the first round. Any number of search steps can be specified on the same command line. For our RsaE motif, let’s try the following 2-step strategy:

```
erpin trset.epn 732bacs.fa 9,17 -umask 17 –nomask
```

In this command, we first unmask element 17, then unmask the whole region. The search time is reduced to 26 minutes and this does not change the number of solutions. To increase search sensitivity and find more distant homologues, we now need to play with scores and pseudocounts. Default cutoff scores are set for each element in order to capture the lowest scoring member of the training set. Pseudocounts are used to allow for additional variation in regions that have as yet displayed very little variation, such as strand 17 in Fig. 7. Playing with scores and pseudocounts can become relatively complex and is beyond the scope of this Chapter. Interested readers should refer to the Erpin documentation. However, the Perl script `erpincommand.pl` provided with the ERPIN distribution can suggest a reasonable ERPIN command using cutoff and pseudocounts to achieve a balanced specificity / sensitivity ratio. The script is run as follows:

```
erpincommand.pl trset.epn
```

and it suggests the following command:

```
erpin trset.epn DATABASE -09,18  -add 09 17 -pcw 51  -cutoff 41.96
```

Running this command (replacing DATABASE by our actual database file name) produces 23 hits after a 176 minutes run. Some hits have poor E-values (1 or more). To filter hits, we can use the script readerpin.pl, as follows, assuming ERPIN output was redirected to `outfile`:

```
readerpin.pl < outfile –e 1
```

This prints out only those hits with an E-value below 1. There are 21 such hits, and the list is the same as before, hence no new significant hit was found. However, one of the two high E-value hits (E = 2.82) is from *Oceanobacillus iheyensis*, a member of Bacilli that was not in the training set and was also found by HMMER (Fig. 8C). This confirms the strength of this candidate.

### *5.6 Searching with INFERNAL*

INFERNAL uses covariance models to describe a structural RNA sequence alignment and subsequently locates instances of this model in genome database. A covariance model is a probabilistic description of the sequence and structure constraints in the training set. It is more general than the model used by ERPIN with the caveat that it cannot take pseudoknot constraints into account. Early versions of INFERNAL required amounts of computing power that kept the program out of reach of most biology labs. However, version 1.0 implemented heuristic filters that increased search speed considerably and rendered INFERNAL searches accessible to anyone with basic command line skills.

First the following command should be executed in order to produce the covariance model from the Stockholm structural alignment `trset.stk` we created above in the HMMER section.

```
cmbuild trset.cm trset.stk
```

The output covariance model is written in file `trset.cm`. In principle, a database scan is possible using this file. However, in order to benefit from E-value calculation and performance-enhancing filters, the model must be processed using the `cmcalibrate` command, as follows:

```
cmcalibrate trset.cm
```

This command is more CPU-intensive than the `cmbuild` command. It requires about 40 minutes in our example, but it can take much longer with complex multi-helix motifs. When it terminates, the cm file is changed to incorporate calibration information. We can now run the search over our 732 genome database using the `cmsearch` command, as follows:

```
cmsearch trset.cm 732bacs.fa
```

As cmsearch is relatively slow, we highly recommend using output redirection and the `nohup` command that enables a program to keep running even if the terminal window is closed:

```
nohup cmsearch trset.cm 732bacs.fa > outfile &
```

Here the `&` sign is used to execute the task in the background. This run takes about 10 hours and finds 26 solutions. In addition to the Firmicute species found by other programs, four identical hits are found in different *Listeria* species (Fig. 9). *Listeria* are Firmicutes and E-values around 0.01 indicate that these are serious candidates. Some *Listeria* hits were found with HMMER but with E-values above 1.

```
>Listeria_monocytogenes_uid43671_CP001602

  Minus strand results:

 Query = 1 - 69, Target = 2392515 - 2392457
 Score = 38.44, E = 0.01941, P = 2.083e-11, GC =  29

          :<<<<<<--<<<<----<<<<<<__________>>>>>>--->>>>-->>>>>>:::::::
        1 UGAAauuAAuCACAAAACAAACAAUACCCCUUUGUUUGAaCGUGaAAaauUUCUCCCAUC 60
          UGAAAU AA CAC AAACAAA:A   CCCCUUU:UUUGA  GUG    AUUUCUCC
  2392515 UGAAAUAAAUCAC-AAACAAAUA--CCCCCUUUAUUUGAUUGUGU-GAAUUUCUCCU--- 2392463

          :::::::::
       61 CCCUUUGUU 69
             UUUG U
  2392462 ---UUUGAU 2392457
```

**Fig. 9.** Extract of an INFERNAL output. Lines 1-4 provide general information about the hit shown below: coordinates of hit in the alignment (query), coordinates of hit in the database sequence (target), score, E-value, P-value and GC-content of hit. The alignment shows the hit sequence (bottom) and the consensus sequence from the training alignment (top). The middle line shows conserved bases. Special characters on top of the alignment show the secondary structure in a bracket notation.

# 6. Conclusion

I have presented five different protocols for RNA motif search, based on BLAST, HMMER, RNAMOTIF, ERPIN and INFERNAL. In absolute terms, the usefulness of an RNA search protocol lies primarily in its precision, or balance between specificity and sensitivity. However practicality is of course important too. In the above examples, the INFERNAL-based protocol was probably the most effective as it found significant candidates that none of the other approaches identified, while enabling reasonable runtimes and command line complexity. However, the diversity of RNA motif search situations is such that alternative protocols should always be considered. Motifs with little or no conservation information are best tackled using RNAMOTIF, motifs based on a pseudoknot are best dealt with ERPIN or RNAMOTIF, and motifs with highly variable or uncertain secondary structures should be addressed preferentially using HMMER or BLAST. The biologist's eye thus remains instrumental in the process of RNA motif search.

# 7. Supplemental data

An archive containing all supplemental files is available at http://rna.igmors.u-psud.fr/suppl_data/

- rsae.fasta: initial RsaE sequence from *S. aureus*.
- trset.fa: training set of RsaE sequences in Fasta format
- trset.stk: training set of RsaE sequences and conserved secondary structure in Stockholm format
- trset.epn: trainings set of RsaE sequences and conserved secondary structure in ERPIN format
- 732bacs.fa: Fasta-formatted database of 732 bacterial genomes (2.5 Gb uncompressed).

# 8. Program Versions and Download Sites

BLAST 2.2.25 from ftp://ftp.ncbi.nlm.nih.gov/blast/executables/blast+/LATEST/

LOCARNA web server: http://rna.informatik.uni-freiburg.de:8080/LocARNA.jsp

HMMER 3.0: http://hmmer.org/

ERPIN 5.4 : http://rna.igmors.u-psud.fr/erpin/

RNAMOTIF: v3.0.7 http://casegroup.rutgers.edu/

INFERNAL : v 1.0.2. http://infernal.janelia.org/

SHUFFLE. Part of SQUID 1.9 library: ftp://selab.janelia.org/pub/software/squid/

# 9. Acknowledgements

I thank Magali Naville and Antonin Marchais for their help in running the different software and Alain Denise for his assistance with Infernal runs and useful comments on the manuscript.